\PassOptionsToPackage{obeyspaces}{url}
\documentclass[sigconf,screen]{acmart}
\usepackage[acm]{definition}

\setcopyright{acmlicensed}
\copyrightyear{2021}
\acmYear{2021}
\acmConference[CIKM '21]{Proceedings of the 30th ACM International Conference on Information and Knowledge Management}{November 1--5, 2021}{Virtual Event, QLD, Australia}
\acmBooktitle{Proceedings of the 30th ACM International Conference on Information and Knowledge Management (CIKM '21), November 1--5, 2021, Virtual Event, QLD, Australia}
\acmPrice{15.00}
\acmDOI{10.1145/3459637.3482088}
\acmISBN{978-1-4503-8446-9/21/11}
\author[Yichen Xu, Yanqiao Zhu, Feng Yu, Qiang Liu, and Shu Wu]{Yichen Xu$^{1,}$*, Yanqiao Zhu$^{2,3,}$*, Feng Yu$^{4}$, Qiang Liu$^{2,3}$, and Shu Wu$^{2,3,5,\dagger}$}

\authornotetext{The first two authors made equal contribution to this work.}
\authornotetext{To whom correspondence should be addressed.}

\affiliation{
	\institution{$^1$School of Computer Science, Beijing University of Posts and Telecommunications}
	\institution{$^2$Center for Research on Intelligent Perception and Computing, Institute of Automation, Chinese Academy of Sciences}
	\institution{$^3$School of Artificial Intelligence, University of Chinese Academy of Sciences \quad $^4$Alibaba Group}
	\institution{$^5$Artificial Intelligence Research, Chinese Academy of Sciences}
	\country{}
}

\email{linyxus@bupt.edu.cn, yanqiao.zhu@cripac.ia.ac.cn, yf271406@alibaba-inc.com, {qiang.liu, shu.wu}@nlpr.ia.ac.cn}

\def\authors{Yichen Xu, Yanqiao Zhu, Feng Yu, Qiang Liu, and Shu Wu}

\begin{document}
\fancyhead{}

\newcommand{\themodel}{\textsf{DESTINE}\xspace}
\xspaceaddexceptions{+}

\title{Disentangled Self-Attentive Neural Networks \mbox{for Click-Through Rate Prediction}}

\begin{abstract}
Click-Through Rate (CTR) prediction, whose aim is to predict the probability of whether a user will click on an item, is an essential task for many online applications.
Due to the nature of data sparsity and high dimensionality of CTR prediction, a key to making effective prediction is to model high-order feature interaction.
%To explicitly model high-order feature interaction, an efficient way is to perform inner product of feature embeddings with self-attentive neural networks.
An efficient way to do this is to perform inner product of feature embeddings with self-attentive neural networks.
To better model complex feature interaction, in this paper we propose a novel DisentanglEd Self-atTentIve NEtwork (\themodel) framework for CTR prediction that explicitly decouples the computation of unary feature importance from pairwise interaction. Specifically, the unary term models the general importance of one feature on all other features, whereas the pairwise interaction term contributes to learning the pure impact for each feature pair.
We conduct extensive experiments using two real-world benchmark datasets. The results show that \themodel not only maintains computational efficiency but achieves consistent improvements over state-of-the-art baselines.
\end{abstract}

\keywords{Click-through rate prediction; high-order feature interaction; disentangled self-attention}

\maketitle

\section{Introduction}

\begin{figure}
	\centering
	\includegraphics[width=\linewidth]{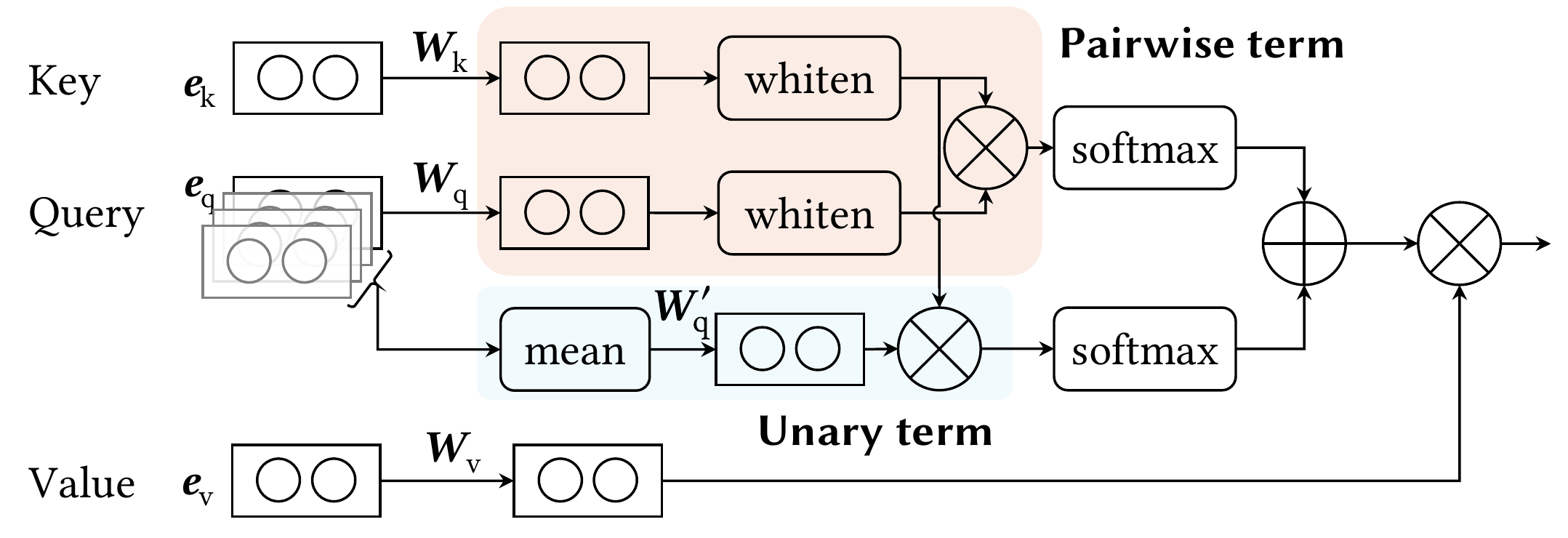}
	\caption{Our proposed disentangled self-attentive networks for CTR prediction that decouple the learning of the pairwise term and the unary term.}
	\label{fig:model}
\end{figure}

Click-Through Rate (CTR) prediction, seeking to predict the probability that a user will interact with a candidate item, is essential for many online applications, such as computational advertising \cite{Liu:2015de} and recommender systems \cite{Cheng:2016jc}.
%For real-world web applications with a large-scale user base, even a slight improvement in prediction accuracy often potentially leads to increased overall revenue. Therefore, considering its high practical values, the CTR prediction problem has attracted a lot of attention from both industry and academia.
One major challenge of making accurate prediction is that the data used in CTR tasks usually involve numerous categorical features, e.g., categories of ads, users' device models, etc., and thus are high-dimensional and extremely sparse, distinct from continuous numerical features such as images.
With such high-dimensional and sparse features as input, one complex model would be inevitably prone to overfitting. Therefore, a successful solution to extracting useful information from these high-dimensional data is to model combinatorial interaction among feature fields based on embedding lookup techniques, also known as \emph{cross features}.
For example, for movie CTR prediction, one informative feature based on third-order feature interaction could be \textsf{\{Age, Gender, Genre\}}, considering that young men tend to prefer action movies.
However, it is not possible to enumerate all combinatorial feature interaction due to the exponential complexity \cite{Cheng:2016jc,Shan:2016cc,Wu:2020eq}. How to automatically model high-order feature interaction thereby attracts a lot of interests.
%Traditionally, handcrafting useful feature interactions is usually done via labor-intensive feature engineering \cite{Richardson:2007eg}.

Recent development in CTR prediction has witnessed a transition from simple linear models \cite{Richardson:2007eg} to more sophisticated methods that model arbitrary-order interaction among these sparse categorical features to make effective prediction.
To model implicit feature interaction, the pioneering work Factorization Machines (FM) \cite{Rendle:2010ja} proposes to model second-order feature interaction via inner-product of embedding vectors.
Following this line, many other methods extend second-order FM to model higher-order interaction \cite{Blondel:2016tj,Guo:2017ja,Lian:2018dl}.
%Following this line, many other methods extend second-order FM to model higher-order interaction, such as HOFM \cite{Blondel:2016tj}, DeepFM \cite{Guo:2017ja}, and xDeepFM \cite{Lian:2018dl}. 
However, these methods suffer from high computational complexity, limiting its practical application in real world.

%In recent years, deep neural network models that are more expressive than linear models have been proposed. In their proposed methods, a multilayer perceptron fed with input features models feature interactions in an implicit manner, leading to suboptimal performance and providing no meaningful explanation of important feature interactions.

One widely-used solution to modeling high-order feature interaction is to compute inner product of feature embeddings \cite{Kang:2018iw,Song:2019iy}, which resembles self-attentive neural networks \cite{Vaswani:2017ul} in deep learning literature.
Specifically, 
%by performing dot product between every feature pair in the embedding space, it allows each feature field to interact with all other fields. 
the dot-product attention scores between every feature pairs can be regarded as the importance of each feature pair. Then, we can compute the second-order feature interaction as weighted sum over feature embeddings. To model arbitrary-order feature interaction, we can stack multiple layers of self-attentive networks with residual connections.

Intuitively, the dot product between each pair of feature embedding vectors encodes \emph{pairwise} semantics of feature interaction, which, however, neglects the modeling of \emph{general} influence of each feature field.
%However, recent literature \cite{Song:2019iy} shows that adding a simple logistic regression (LR) network based on feature embeddings, i.e. first-order feature interaction, helps boost the performance. The LR module is intended to directly model importance of each feature field.
%Moreover, as suggested in \citet{Song:2019iy}, the attentive scores of feature interaction could be interpreted from another aspect, i.e. the general influence of each feature field, which can be obtained by taking average over attention scores. This \emph{general} influence models the individual influence of each feature, which carries \emph{unary} semantics.
%{\color{red}Based on this observation, a natural question to ask is \textit{can we simply model both the pairwise and unary semantics using dot-product-based self-attentive networks}?}
%Based on this observation, we dig deep into the formulation of the self-attention module by dissecting it into two terms,
To explicitly model such \emph{unary} semantics, we propose to decouple a unary term from the vanilla self-attention network that computes the \emph{general} impact of one certain feature to all other features.
We term the resulting framework \ul{D}isentangl\ul{E}d \ul{S}elf-at\ul{T}ent\ul{I}ve \ul{NE}twork, \themodel for brevity.
In particular, to better model feature interaction, \themodel consists of two independent computational blocks illustrated in Figure \ref{fig:model}: a \emph{whitened} pairwise term that models \emph{specific} interaction between two features and a unary term for the \emph{general} influence of one feature to all others. %We find that the dot-product-based self-attention network implicitly models both terms but the two terms are tightly interrelated in the learning process.

For the CTR prediction problem, we first embed input features into low-dimensional spaces and then compute high-order feature interaction by stacking multiple disentangled self-attentive layers. Finally, the embeddings resulting from the last interaction layer are used to estimate the click behavior.
Extensive experiments on two real-world datasets demonstrate that our proposed \themodel not only achieves state-of-the-art performance but also retains high computational efficiency.
Our code is made publicly available at \url{https://github.com/CRIPAC-DIG/DESTINE}.

\section{The Proposed \themodel Approach}

\subsection{Problem Definition}

Suppose the training dataset \(\mathbb{D} = \left\{\bm{x}_i, y_i\right\}_{i = 1}^N\) contains \(N\) samples, where each sample \(\bm{x}_i\) consists of \(M\) fields of users' and items' features and its associated label \(y_i \in \{0, 1\}\) represents that user's behavior (e.g., whether to click an item).
The problem of click-through rate prediction is to predict \(\hat{y}_i\), given a feature vector \(\bm{x}_i\), for accurately estimating whether a user will interact with an item.

\subsection{Learning Decoupled Feature Interaction}
\label{sec:self-attention-module}

The proposed \themodel consists of three key components: (a) the embedding layer, (b) the interaction layer, and (c) the output layer.
At first, the input features are fed into the embedding layer, which transforms input features into dense, low-dimensional embedding vectors.
Then, these feature embeddings are fed into several stacked interaction layers, which model high-order interaction.
After that, we feed the embeddings from the last interaction layer into the output layer to estimate the click behavior.

For each sparse input feature \(\bm{x}_i\), we transform it into dense embeddings \(\bm{e}_i \in \mathbb{R}^d\) via embedding lookup. Once we obtained a compact representation for each feature, we use the scaled dot-product attention scheme to model high-order feature interaction among feature fields.
Specifically, we formulate each feature interaction as a (\textit{key}, \textit{value}) pair and learn the importance of each feature interaction by multiplying each feature embedding, such that important key--value pairs get higher attention scores.
Formally, we first transform each feature embedding into a new embedding space \(\mathbb{R}^{d'}\) as follows
%\begin{minipage}{.5\linewidth}
	\begin{equation}
		\bm{q}_m = \bm{W}_\text{q} \bm{e}_m,
	\end{equation}
%\end{minipage}
%\begin{minipage}{.5\linewidth}
	\begin{equation}
		\bm{k}_n = \bm{W}_\text{k} \bm{e}_n,
	\end{equation}
%\end{minipage}
where the \emph{query} and \emph{key} transformation are parameterized by two linear transformation matrices \(\bm{W}_\text{q}, \bm{W}_\text{k} \in \mathbb{R}^{d' \times d}\), respectively.

Then, we compute the correlation \(\alpha(\bm{e}_m, \bm{e}_n)\) between feature \(m\) and feature \(n\).
Previous work \cite{Yin:2020wb} in visual representation learning demonstrates that the importance score of feature \(m\) over feature \(n\) could be decomposed into two terms: a pairwise term to model pure specific interaction and a unary term to model general impact over all feature fields.
We take summation of these two terms:
\begin{equation}
	\alpha(\bm{e}_m, \bm{e}_n) = \alpha_\text{p}(\bm{e}_m, \bm{e}_n) + \alpha_\text{u}(\bm{e}_m, \bm{e}_n).
	\label{eq:disentangled-self-attention}
\end{equation}

For the pairwise term, we perform whitening \cite{Friedman:1987ho} on the key and the query vector to model pure interaction among features, which makes the two interacting features less correlated with each other:
\begin{equation}
	\alpha_\text{p}(\bm{e}_m, \bm{e}_n) = \sigma\left(\left(\bm{q}_m - \bm\mu_\text{q}\right)^\top\left(\bm{k}_n - \bm\mu_\text{k}\right)\right),
	\label{eq:pairwise}
\end{equation}
where \(\sigma(\cdot)\) is the softmax function; \(\bm\mu_\text{q} = \frac{1}{M} \sum_{i = 1}^M \bm{W}_\text{q} \bm{e}_i\) and \(\bm\mu_\text{k} = \frac{1}{M} \sum_{j = 1}^M \bm{W}_\text{k} \bm{e}_j\) takes average of the key and the query vectors, respectively.

Regarding the unary term, we introduce another query transformation matrix \(\bm{W}_\text{q}' \in \mathbb{R}^{d' \times d}\) for modeling significant features:
\begin{equation}
	\alpha_\text{u}(\bm{e}_m, \bm{e}_n) = \sigma\left((\bm\mu_\text{q}')^\top \bm{k}_n \right),
	\label{eq:unary}
\end{equation}
where \(\bm\mu_\text{q}'\) is the mean vector from another transformation to model the general impact on key vectors, i.e.
\(
	\bm\mu_\text{q}' = \frac{1}{M} \sum_{i = 1}^M \bm{W}_\text{q}' \bm{e}_i.
\)

After computing the attention score for each feature interaction \((m, n)\), we transform each candidate feature to a new embedding space by a \emph{value} transformation, parameterized by a linear transformation matrix \(\bm{W}_\text{v} \in \mathbb{R}^{d' \times d}\), as given in the sequel,
\begin{equation}
	\bm{v}_k = \bm{W}_\text{v} \bm{e}_k.
\end{equation}
At last, we update the final representation for feature field \(m\) by linearly combining all features.

\paragraph{Extension to multihead self-attention.}
To allow the model to learn distinct feature interaction in different subspaces, we use multiple attention heads.
Specifically, we compute the representation of feature field \(m\) under an attention head \(h\) by
\begin{equation}
	\bm{z}_{m}^{(h)} = \sum_{k = 1}^M \alpha^{(h)}(\bm{e}_m, \bm{e}_k) \cdot \bm{v}_k^{(h)},
\end{equation}
where \(\alpha^{(h)}\) is the attention score computed via Eq. (\ref{eq:disentangled-self-attention}). Note that, each attention head \(h\) keeps its distinct weight parameters \(\bm{W}_\text{k}^{(h)}\), \(\bm{W}_\text{q}^{(h)}\), \((\bm{W}_\text{q}^\prime)^{(h)},\) and \(\bm{W}_\text{v}^{(h)}\).

Then, we obtain $z_m$, the overall hidden representation for feature $m$, by concatenating the representation of all attention heads as
\begin{equation}
	\bm{z}_m = \left[\bm{z}_m^{(1)}; \bm{z}_m^{(2)}; \dots; \bm{z}_m^{(H)}\right],
\end{equation}
where \(H\) is the number of attention heads.
Additionally, following previous work \cite{Song:2019iy,Yu:2020ww}, we incorporate raw, individual features (first-order features) via residual connections \cite{He:2016ib}, which is formulated as,
\begin{equation}
	\hat{\bm{z}}_m = \varphi(\bm{z}_m + \bm{W}_\text{r} \bm{e}_m),
	\label{eq:interaction-output}
\end{equation}
where \(\bm{W}_\text{r} \in \mathbb{R}^{d'H \times d}\) is a linear projection matrix to avoid dimension mismatch and \(\varphi(\cdot) = \max(0, \cdot)\) is the ReLU activation function.

With features \(\hat{\bm{z}}_m\) received from the last interaction layer, we concatenate all \(M\) features and utilize a simple logistic regression model on top of them to predict user behavior, formulated as
\begin{equation}
	\hat{y} = \sigma\left(\bm{w}^\top \left[\hat{\bm{z}}_1; \hat{\bm{z}}_2; \dots ; \hat{\bm{z}}_M\right] + b\right),
\end{equation}
where \(\bm{w} \in \mathbb{R}^{d'HM}\) is a linear projection vector, \(b\) is a bias term, and \(\sigma(x) = 1 / (1 + e^{-x})\) is the sigmoid function.

At last, we employ the binary cross-entropy function, which is widely-used in CTR prediction models, as the loss function,
\begin{equation}
	\mathcal{L} = -\frac{1}{N} \sum_{(y, \hat{y}) \in \mathbb{D}} \left(y \log \hat{y} + (1 - y) \log (1 - \hat{y}) \right), \label{eq:loss}
\end{equation}
where \(y\) and \(\hat{y}\) is the ground truth and the predicted click, respectively.
We use gradient descent algorithms to update model weights.

\subsection{Complexity Analysis}

%For the proposed DSAN model, we only introduce a new learnable parameter \(\bm{W}_\text{m}\) for each attention head, leading to additional space complexity of \(O(d'H)\) for each interaction layer. In addition, the time complexity of DSAN is \(O(H(d'dM + d'M^2 + d'M2)) = O(MHd' (2M+d))\), compared to that of the original self-attention module of \(O(MHd' (M+d))\). Therefore, the modification to the original self-attention module results in negligible computational overhead. Also, note that in our implementation, \(d, d'\), and \(H\) are usually small, which demonstrates that our model is memory-efficient and computation-friendly.

For the proposed model, we only introduce a new learnable parameter \(\bm{W}_\text{q}^\prime \in \mathbb R^{d \times d^\prime}\) for each attention head, leading to additional space complexity of \(O(dd'H)\) for each interaction layer. In addition, the time complexity of \themodel is \(O(H(d' d M + d^\prime M + d' M^2 + d' M^2)) = O(MHd' (2M + d + 1))\), compared to that of the original self-attention module of \(O(MHd'(M + d))\). Note that \(d, d'\), and \(H\) are usually small, demonstrating that our model is memory-efficient and computation-friendly.

%\subsubsection{Remarks on modeling high-order feature interaction}
%Finally, we analyze the ability of modeling arbitrary-order combinatorial features of the proposed \themodel model.
%Suppose that we stack \(L\) interaction layers in our model, with residual connections between two consecutive layers.
%In the first interaction layer, each individual feature (i.e. the first-order feature) interacts with all other features; their relationship can be captured by self-attention scores. Therefore, all second-order combinatorial features can be encoded in the output of the first interaction layer, where the non-addition property is ensured by non-linear activation functions in Eq. (\ref{eq:interaction-output}).
%Then, with the help of intra-layer residual connections and the disentangled unary term, we also incorporate first-order features \(\bm{e}_1\) to \(\bm{e}_M\) into the final representation \(\hat{\bm{z}}_m\) for feature \(m\).
%
%Given that the first interaction layer is able to encode first- and second-order feature interaction, we can stack many interaction layers as the order of feature interaction goes higher. For example, forth-order feature interaction can be obtained by combining outputs from the first interaction layer.
%Our proposed \themodel model is thereby able to model arbitrarily high-order feature interaction. However, note that in practice, features interaction greater than fifth-order are rarely used \cite{Blondel:2016tj, Lian:2018dl, Yu:2020ww}. Therefore, it is not necessary to stack too many interaction layers.

\section{Experiments}
In this section, we present empirical analysis to answer the following three questions.

\textbf{RQ1}. Does the proposed \themodel method outperform existing state-of-the-art baseline methods that model feature interaction for CTR prediction?

\textbf{RQ2}. Many models integrate implicit feature interaction via deep neural networks (DNN); how does the proposed model with implicit feature interaction compare with them?

\textbf{RQ3}. How do other model variants perform compared to the proposed \themodel?

\subsection{Experimental Setup}

We use two large-scale datasets Avazu and Criteo for evaluation.
\textbf{Avazu} contains click logs of mobile ads in 10 days, which is composed of 23 categorical features including domains, categories, connection types, etc.
\textbf{Criteo} comprises traffic logs of display ads over 7 days. Each sample contains 13 numerical and 26 categorical feature fields.
The detailed statistics is summarized in Table \ref{tab:dataset-statistics}.
%For preprocessing the two datasets, we closely follow the settings in \citet{Zhu:2020wn}.

For fair comparison, following most existing studies \cite{Song:2019iy,Li:2019jm,Zhu:2020wn}, we randomly sample 80\% data as the training set, 10\% as the test set, and the remaining 10\% as the validation set.
We report the performance in terms of two widely-used metrics AUC and logloss.
Please kindly note that considering a large scale of user base, performance improvements of AUC at \emph{1\textperthousand-level} are considered as practically significant for industrial deployment  \cite{Cheng:2016jc,Guo:2017ja,Wang:2017gy,Song:2019iy,Zhu:2020wn,Wu:2020eq}.

\begin{table}[t]
	\small
	\caption{Dataset statistics}
	\label{tab:dataset-statistics}
	\begin{tabular}{ccccc}
		\toprule
		Dataset & \# Instances & \# Fields & \# Features & Positives \\
		\midrule
		Avazu & 40,428,967 & 23 & 1,544,488 & 17\% \\
		Criteo & 45,840,617 & 39 & 998,960 & 26\% \\
		\bottomrule
	\end{tabular}
\end{table}

\begin{table}[t]
	\caption{Performance (AUC and logloss) and training time (mins) of methods that model feature interaction. The best performance is highlighted in boldface.}
	\label{tab:results-feature-interaction}
	\begin{adjustbox}{max width=\linewidth}
	\begin{threeparttable}
	\begin{tabular}{lccccccccc}
	\toprule
	\multirow{2.5}[0]{*}{Model} & \multicolumn{3}{c}{Criteo} & \multicolumn{3}{c}{Avazu} \\
	\cmidrule(lr){2-4} \cmidrule(lr){5-7}
			& AUC   & Logloss & Time & AUC   & Logloss & Time \\
    \midrule
	\rowcolor{gray!10}
	LR    & 0.7820 & 0.4695 & 535.2 & 0.7560 & 0.3964 & 342.6 \\
	
	FM    & 0.7836 & 0.4700 & 391.3 & 0.7706 & 0.3856 & 480.2 \\
	AFM   & 0.7938 & 0.4584 & 468.3 & 0.7718 & 0.3854 & 130.7 \\
	
	\rowcolor{gray!10}
	DeepCrossing & 0.8009 & 0.4513 & --- & 0.7643 & 0.3889 & --- \\
	\rowcolor{gray!10}
	CrossNet & 0.7907 & 0.4591 & 216.7 & 0.7667 & 0.3868 & 56.3 \\
	\rowcolor{gray!10}
    CIN   & 0.8009 & 0.4517 & 219.0 & 0.7758 & 0.3829 & 179.6 \\
	\rowcolor{gray!10}
    HOFM  & 0.8005 & 0.4508 & 696.2 & 0.7701 & 0.3854 & 903.0 \\
	\rowcolor{gray!10}
    AutoInt & 0.8061 & 0.4455 & 375.9 & 0.7752 & 0.3824 & 112.6 \\
	\rowcolor{gray!10}
    \themodel  & \textbf{0.8087} & \textbf{0.4425} & 477.3 & \textbf{0.7831} & \textbf{0.3789} & 104.9 \\
	\bottomrule
	\end{tabular}
	\begin{tablenotes}
	\scriptsize{
	\note For fair comparison, the training time of DeepCrossing is not listed since its implementation is based on distributed multi-GPU environments.
	}
	\end{tablenotes}
	\end{threeparttable}
	\end{adjustbox}
\end{table}

\subsection{Performance of Feature Interaction Models}
\label{sec:performance-comparison}

\subsubsection{Baselines.}
Representative baseline CTR prediction models can be grouped into three lines, according to the order of feature interaction they model: (a) first-order method LR \cite{Richardson:2007eg}, (b) second-order methods FM \cite{Rendle:2010ja} and AFM \cite{Xiao:2017et}, and (c) higher-order methods DeepCrossing \cite{Shan:2016cc}, CrossNet \cite{Wang:2017gy}, CIN \cite{Lian:2018dl}, HOFM \cite{Blondel:2016tj}, and AutoInt \cite{Song:2019iy}.
In this section, we only include counterparts that directly model feature interaction; for full models that involve a DNN component to model implicit interaction, such as Deep\&Cross and AutoInt+, the results are presented later in Section \ref{sec:implicit-interaction-comparison}.

\subsubsection{Implementation details.}

We set the number of attention heads to \(H = 2\); the hidden dimension of embeddings is set to \(d' = 32\) and the size of attentive embeddings is set to \(d = 64\).
The model is trained using the Adam optimizer \cite{Kingma:2015us} with a learning rate of \(0.001\). We find that the weight for \(\ell_2\) regularization slightly impacts the model performance, and thus we carefully tuned it in \([5\times 10^{-3}, 5\times 10^{-4}, 5\times 10^{-5}, 5\times 10^{-6}]\). Moreover, we set the dropout \cite{Srivastava:2014cg} rate to \(0.2\) to avoid overfitting.

\subsubsection{Results and analysis.}

The overall performance is summarized in Table \ref{tab:results-feature-interaction}. We also report the total training time (mins) of each model. All baseline performance is referenced from their original papers. The training time of baselines is measured using their official implementations.
Overall, from the table, it is evident that the proposed \themodel achieves the best performance on all datasets. Moreover, \themodel also enjoys another merit of low computational complexity as it takes relatively low time to finish training.

We also make other observations.
Firstly, methods that model higher-order feature interaction generally gain larger performance improvements, which verifies the necessity of modeling high-order interaction in CTR prediction.
Secondly, although several high-order methods such as DeepCrossing utilize a feedforward neural network to learn feature interaction, their performance is even inferior to second-order methods, which implies that it is not sufficient to learn useful feature interaction in an implicit way. Our method, on the contrary, explicitly models useful feature interaction via attentive networks, so that achieves better performance.
Thirdly, compared to AutoInt that leverages a vanilla attentive net, our proposed \themodel disentangles a pairwise and a unary term from computing attention scores, which further boosts performance.
In summary, the results validate the effectiveness of \themodel.

\subsection{Performance of Deep Models Integrated}
\label{sec:implicit-interaction-comparison}

\subsubsection{Evaluation protocols.}
Many existing CTR prediction models alternatively integrate a Deep Neural Network (DNN) component to learn implicit feature interaction.
For each DNN layer, we use a simple feedforward network, followed by a batch normalization layer \cite{Ioffe:2015ud}. %, as given by,
%\begin{equation}
%	\widetilde{\bm{z}}_m = \operatorname{BN}\left(\operatorname{ReLU}\left(\bm{W}_\text{DNN} \bm{e}_m^\top + \bm{b}_\text{DNN}\right)\right),
%\end{equation}
%where \(\bm{W}_\text{DNN}\) and \(\bm{b}_\text{DNN}\) are weight parameters for each DNN layer.
We stack two DNN layers for all datasets. The resulting representation \(\widetilde{\bm{z}}_m\) for feature \(m\) is concatenated with \(\hat{\bm{z}}_m\) and is further trained with another linear layer to obtain the final embedding. %, which enables joint training of the DNN component and the disentangled self-attentive network.
For investigating whether implicit feature interaction improves the performance, we include the following baselines: Wide\&Deep \cite{Cheng:2016jc}, DeepFM \cite{Guo:2017ja}, Deep\&Cross \cite{Wang:2017gy}, xDeepFM \cite{Lian:2018dl}, AutoInt+ \cite{Song:2019iy}, DeepIM \cite{Yu:2020ww}, and AutoCTR \cite{Song:2020do}.
Following the naming convention, we term our hybrid model \themodel+ that jointly learns feature interaction using decoupled attentive networks and DNNs.

\subsubsection{Results and analysis}

The results are summarized in Table \ref{tab:results-implicit-interaction}.
We observe that \themodel+ achieves new state-of-the-art performance over existing baselines, which demonstrates that integrating with implicit feature interaction with DNN, \themodel is able to learn feature interaction more effectively.
Moreover, from the last two columns, we observe that the average improvements brought by the DNN component are limited. This indicates that the base model \themodel has already achieved relatively high performance, which once again verifies the efficacy of \themodel.
We also note that the relative improvements on the Criteo dataset are greater than that on the Avazu. This could be explained by that the number of feature fields on Criteo is much more than that on Avazu; in other words, Criteo is ``wider'' than Avazu, where DNN training is much easier than the proposed disentangled self-attentive model.

\begin{table}[t]
	\centering
	\caption{Performance of models integrated with DNNs that implicitly model feature interaction. The averaged changes in the last column are relative changes of performance compared to the corresponding base models.}
	\resizebox{\linewidth}{!}{
	\begin{tabular}{lcccccc}
	\toprule
	\multirow{2.5}[0]{*}{Model} & \multicolumn{2}{c}{Criteo} & \multicolumn{2}{c}{Avazu} & \multicolumn{2}{c}{Avg. Changes} \\
	\cmidrule(lr){2-3} \cmidrule(lr){4-5} \cmidrule(lr){6-7}
       & AUC   & Logloss & AUC   & Logloss & AUC   & Logloss \\
	\midrule
    Wide\&Deep & 0.8026 & 0.4494 & 0.7749 & 0.3824 & \(+0.0292\) & \(-0.0213\) \\
    DeepFM & 0.8066 & 0.4449 & 0.7751 & 0.3829 & \(+0.0142\) & \(-0.0113\) \\
    Deep\&Cross & 0.8067 & 0.4447 & 0.7731 & 0.3836 & \(+0.0200\) & \(-0.0164\) \\
    xDeepFM & 0.8070 & 0.4447 & 0.7770 & 0.3823 & \(+0.0068\) & \(-0.0096\) \\
    AutoInt+ & 0.8083 & 0.4434 & 0.7774 & 0.3811 & \(+0.0023\) & \(-0.0020\) \\
    DeepIM & 0.8044 & 0.4472 & 0.7828 & 0.3809 & \(+0.0165\) & \(-0.0138\) \\
    AutoCTR & 0.8104 & 0.4413 & 0.7791 & 0.3800 & ---   & --- \\
    \themodel+ & \textbf{0.8118} & \textbf{0.4398} & \textbf{0.7851} & \textbf{0.3779} & \(+0.0026\) & \(-0.0019\) \\
	\bottomrule
	\end{tabular}
	}
	\label{tab:results-implicit-interaction}
\end{table}

\begin{figure}
	\centering
	\includegraphics[width=\linewidth]{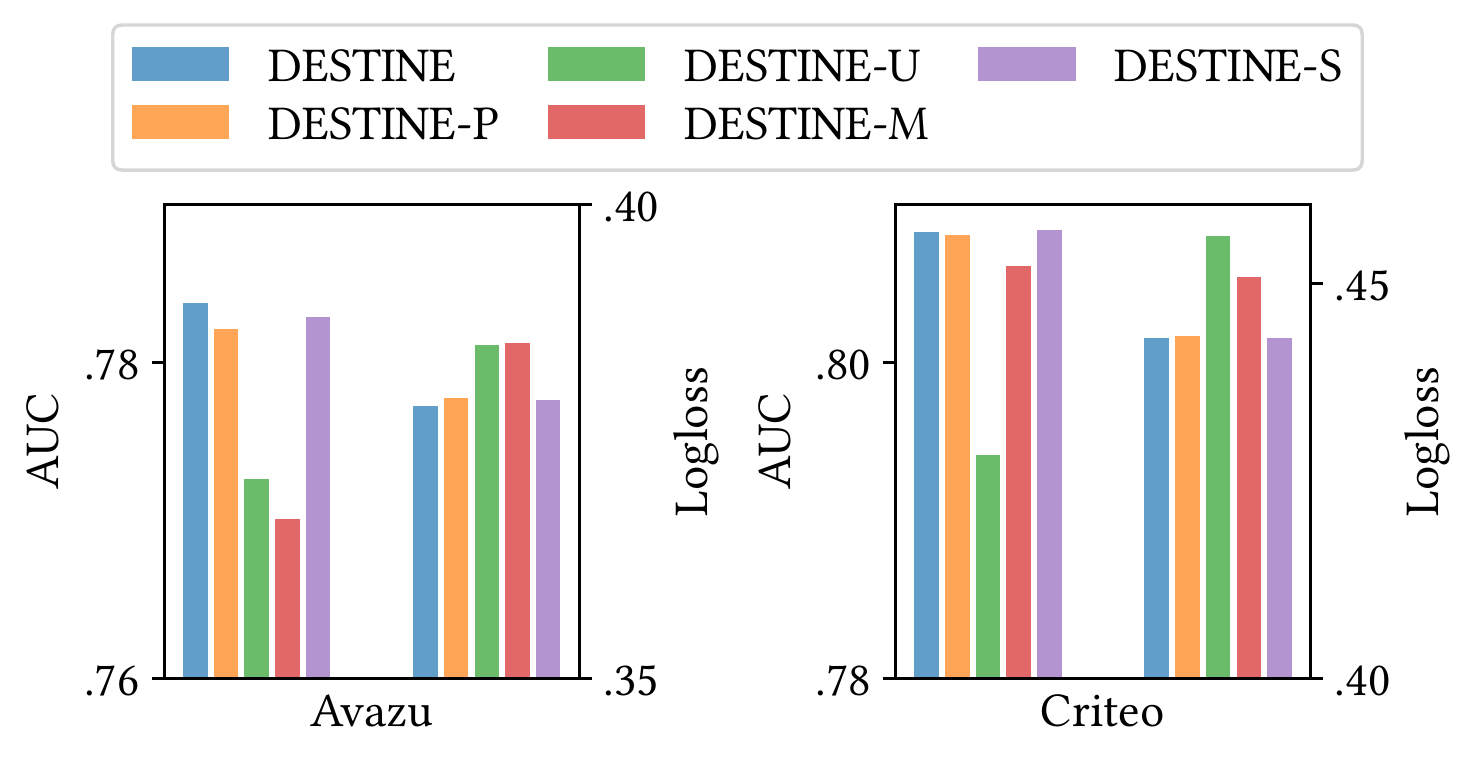}
	\caption{Performance of ablations and other model variants.}
	\label{fig:ablation}
\end{figure}

\subsection{Performance of Model Variants}

For diagnosing the proposed scheme, we further investigate four variants of the proposed disentangled self-attention module:
\begin{itemize}
	\item \textbf{Pairwise} (\themodel-P). We only use the pairwise term in Eq. (\ref{eq:pairwise}).
	\item \textbf{Unary} (\themodel-U). We only use the unary term in Eq. (\ref{eq:unary}).
	\item \textbf{Multiplication} (\themodel-M). We use multiplication rather than addition for combining the pairwise and the unary term, as described below,
	\begin{equation}
		\alpha_\text{d}(\bm{e}_m, \bm{e}_n) = \sigma\left(\left(\bm{q}_m - \bm\mu_\text{q}\right)^\top\left(\bm{k}_n - \bm\mu_\text{k}\right)\right) \cdot \sigma\left((\bm\mu_\text{q}')^\top \bm{k}_n\right).
	\end{equation}
	\item \textbf{Shared transformation} (\themodel-S). We consider preserving the shared key transformation \(\bm{W}_\text{k}\) in the unary term, as formulated below,
	\begin{equation}
		\alpha_\text{s}(\bm{e}_m, \bm{e}_n) = \sigma\left(\left(\bm{q}_m - \bm\mu_\text{q}\right)^\top\left(\bm{k}_n - \bm\mu_\text{k}\right)\right) + \sigma\left(\bm\mu_\text{q}^\top \bm{k}_n\right).
	\end{equation}
\end{itemize}

The results are summarized in Figure \ref{fig:ablation}. We can see that both the pairwise and the unary term benefits model performance. However, the way to combine the two terms plays an important role. \themodel-M performs poorly despite its utilization of both terms and is even outperformed by \themodel-U and \themodel-P, where only one of the two terms is used. This verifies our justification that coupled gradient in \themodel-M will deteriorate the performance. The performance \themodel-S is also inferior to \themodel, since the shared \(\bm{W}_\text{q}\) transformation still leads to the coupling of gradients during learning. The outstanding performance of \themodel compared to all other variants justifies the design of our model.

\section{Conclusion}
In this paper, we present a disentangled self-attention network \themodel for click-through rate prediction, which consists of two terms for pairwise and unary semantics. Specifically, the unary term models the general impact of one feature on all others, whereas the remaining whitened pairwise term models pure feature interaction among each feature pair. Extensive experiments on two real-world datasets demonstrate the effectiveness of the proposed method.

\begin{acks}
This work is supported by National Key Research and Development Program (2018YFB1402600), National Natural Science Foundation of China (61772528), and Shandong Provincial Key Research and Development Program (2019JZZY010119).
\end{acks}

\bibliographystyle{ACM-Reference-Format}
\bibliography{cikm2021}

\end{document}